\documentclass[useAMS,usenatbib]{mn2e}
\pdfoutput=1
\usepackage{graphicx}
\usepackage{eufrak}
\usepackage{slashed}
\usepackage{bm}
\usepackage{amsmath}
\usepackage{epsfig}
\newcommand{\exclude}[1]{}

\title{Large Pulsar Kicks from Topological Currents}
\author[James Charbonneau, Kelsey Hoffman and Jeremy Heyl]{James Charbonneau$^{1}$
Kelsey Hoffman$^{1}$
and Jeremy Heyl$^{1}$\thanks{Canada Research Chair}\\
$^1$Department of Physics \& Astronomy, University of British
Columbia, 6224 Agricultural Road, Vancouver, V6T 1Z1, Canada;\\
E-mail:
james@phas.ubc.ca, kelsey@phas.ubc.ca and heyl@phas.ubc.ca}

\begin{document}

\date{Accepted ---. Received ---; in original form ---}
\date{}

\pagerange{\pageref{firstpage}--\pageref{lastpage}} \pubyear{2009}

\maketitle

\label{firstpage}

\begin{abstract}
We show that kicks generated by topological currents may be responsible for the large velocities seen in a number of pulsars. The majority of the kick builds up within the first second of the star's birth and generates a force about two orders of magnitude larger than a neutrino kick in the same temperature and magnetic field regime. Because of the nature of the topological currents the star's cooling is not affected until it reaches $10^9$~K; thereafter the current replaces neutrino emission as the dominant cooling process. A requirement for the kick to occur is a suitably thin crust on the star; this leads us to speculate that pulsars with large kicks are quark stars and those with small kicks are neutron stars. If true this would be an elegant way to distinguish quark stars from neutron stars.
\end{abstract}


\section{Introduction}

The goal of this paper is to elaborate on a kick mechanism first discussed by \citet{Charbonneau:2009ax} that may explain pulsar velocities greater than $1000$~km$\,$s$^{-1}$. There have been a number of studies that have compiled and modelled the velocities of pulsars. Although they disagree on whether the distribution is indeed bimodal, they agree that a significant number of pulsars are travelling faster than can be attributed to neutrino kicks. The analysis of \citet{Arzoumanian:2001dv} favours a bimodal velocity distribution with peaks at 90~km$\,$s$^{-1}$ and 500~km$\,$s$^{-1}$ with $15\%$ of pulsars travelling at speeds greater than 1000 km$\,$s$^{-1}$. Alternatively \citet{Hobbs:2005yx} and \cite{0004-637X-643-1-332} both predict a single peaked distribution with an average velocity of $\sim 400$~km$\,$s$^{-1}$, but point out that the faster pulsars B2011+38 and B2224+64 have speeds of $\sim 1600$~km$\,$s$^{-1}$. Large velocities are unambiguously confirmed with the model independent measurement of pulsar B1508+55 moving at $1083^{+103}_{-90}$~km$\,$s$^{-1}$ \citep{Chatterjee:2005mj}.

Currently no mechanism exists that can reliably kick the star hard enough to reach these velocities.  Asymmetric explosions can only reach 200~km$\,$s$^{-1}$ \citep{Fryer:2003tc}, and asymmetric neutrino emission is plagued by the problem that at temperatures high enough to produce the kick the neutrinos are trapped inside the star \citep{Sagert:2007ug}. Alterations of the neutrino model that take into account only a thin shell of neutrinos require large temperatures and huge surface magnetic fields.


The kick mechanism we will discuss relies on the existence of topological vector currents of the form described by \citet{Charbonneau:2009ax},
\begin{eqnarray}  
\label{j}
\langle j \rangle = (\mu_L - \mu_R)\frac{e \Phi}{2\pi^2},
\end{eqnarray}  
where $\mu_R$ and $\mu_L$ are the chemical potential of the right and left-handed electrons, and $\Phi$ is the magnetic flux.   There are three requirements for topological vector currents to be present: an imbalance in left and right-handed particles $\mu_L\neq \mu_R$, degenerate matter $\mu\gg T$, and the presence of a background magnetic field $B\neq 0$.  All of these are present in neutron and quark stars.  The weak interaction, by which the star attains equilibrium, violates parity; particles created in this environment are primarily left-handed.  The interior of the star is very dense, $\mu_e \sim 100$~MeV, and cold, $T\sim1$~MeV, such that the degeneracy condition $\mu\gg T$ is met, and neutron stars are known to have huge surface magnetic fields, $B_\textrm{s}\sim10^{12}$~G. Interested readers are encouraged to read \citet{Charbonneau:2009ax}.

Kicks caused by the topological current are very similar to neutrino kicks. The helicity of the electrons and the single spin state in the lowest Landau level allow electrons to travel with a preferred direction through the neutron star. The difference between the two mechanisms is the linear momentum that the particle transfers to the star. Neutrinos have an energy close to the temperature of the star, while the electrons, which are degenerate, have an energy close to their Fermi momentum. We will discuss this critical difference in Section~\ref{difference}.

If the electrons carried by the current can transfer their momentum into space---either by being ejected or by radiating photons---the current could push the star like a rocket.  In typical neutron stars this is unlikely because the crust (the region where $\mu\sim T$) is thought to be about 1~km thick. Once it reaches this thick crust, the current will likely travel along it and be reabsorbed into the bulk of the star. But if the crust is very thin, the electrons may leave the system or emit photons that will carry their momentum to space. The crust for bare quark stars is thought to be about 1000~fm \citep[][see also recent developments by \cite{Alford:2006bx}]{Alcock:1986hz}. With this in mind we conjecture that stars with very large kicks, $v\gg 200$~km$\,$s$^{-1}$, are quark stars and that slow moving stars, $v \leq 200$~km$\,$s$^{-1}$, are kicked by some other means, such as asymmetric explosions or neutrino emission, and are typical neutron stars.  Confirmation of this would provide an elegant way to discriminate between neutron stars and quark stars.

 \section{Simple model of quark stars}
The kick will likely only occur in quark stars that are bare or have very thin crusts. We will consider only the existence of light quarks in the star, which attain equilibrium through the quark direct Urca processes,  
\begin{eqnarray}  
\label{eq:quark_urca}
e^- + u\ (+\ \bar{\nu}_e) &\leftrightarrow& d\ (+\ \nu_e) \\ 
e^- + u\ (+\ \bar{\nu}_e) &\leftrightarrow& s\ (+\ \nu_e)\,,
\end{eqnarray}
where the neutrino terms only appear in the final state of the interaction. The direct Urca processes in quark matter do not have the same Fermi momentum restrictions as in neutron matter and are thus not suppressed. These processes are at equilibrium when, 
\begin{equation}    
\mu_u + \mu_e = \mu_d ~\mathrm{and}~ \mu_u + \mu_e = \mu_s\,. 
\end{equation}
In this balance we ignore the neutrino chemical potential. In discussing the kick we will focus on temperatures where the star makes the transition from being opaque to neutrinos to being transparent. We will refer to this temperature range as the translucent regime. In the opaque regime the neutrino chemical potential is nonzero. As the star cools the mean free path of the neutrinos becomes large enough that all neutrinos created may freely leave the star. This is precisely the temperature at which the topological current begins to flow and creates a kick. In discussing these temperatures there is the concern that the neutrino chemical potential cannot be ignored. However, once in the transparent regime the neutrino chemical potential dissipates much faster than the time scale of the kick, thus we can safely ignore it.

The quark matter must also be electrically neutral,
\begin{eqnarray}    
\label{eq:charge_balance}
\rho/e = \frac{2}{3}n_u - \frac{2}{3}n_d - \frac{1}{3}n_s - n_e = 0\,,
\end{eqnarray}
where $\rho$ is the charge density and $n_i$ are the number densities of the quarks and electrons. We will consider the simplest model where the quark masses are set to zero, thus their Fermi momenta are equal to their Fermi energies.  Unfortunately, this approximation requires the electron density to be zero to satisfy equation (\ref{eq:charge_balance}), thus removing an essential element from the star: electrons. The strange quark mass is quite large and the number density should be described using nonrelativistic Fermi statistics, which readmits elections into the mix to conserve charge. We will follow \citet{Iwamoto:1982} by assuming that the quarks are all massless and an electron chemical potential is present to capture the contribution of the mass of the strange quark. The Fermi momentum of the quark species can be written in terms of the baryon number density, denoted $n_b$, as

\begin{eqnarray}     
\label{eq:k_quark}
k_q = (\pi^2 n_b)^{1/3} \sim 235 \left( \frac{n_b}{n_0}\right)^{1/3} \textrm{ MeV}\,.
\end{eqnarray}
For typical densities in the core of the neutron star the Fermi momentum is $k_q \sim 400$ MeV, which is greater than the strange quark mass. Neglecting the mass of the strange quark is a crude approximation but works well for this calculation. We can obtain the electron Fermi momentum using the fraction of electrons to baryons, $Y_e = n_e/n_b$, which yields
\begin{eqnarray}     
\label{eqn:k_e}
k_e  = (3Y_e)^{1/3}k_q \sim 73  \left( \frac{n_b}{n_0}\right)^{1/3}  \textrm{ MeV}\,.
\end{eqnarray}
Because $k_e \gg m_e$, we find that $k_e \sim \mu_e$. The typical value for the electron fraction is $Y_e = 0.01$. This value is not critical in determining the strength of the kick because we will find that the dependance on $k_e$ cancels. However, as we will discuss in Section~\ref{difference}, the value of the Fermi momentum $k_e$ when compared to the core temperature of the star $T_\textrm{core}$ is critical in understanding how topological currents generate large kicks.
 
\section{Cooling of quark stars}

In order to properly determine the size of a kick we need to understand how the core temperature of the quark star evolves with time. Unfortunately, kicks are likely to occur right after the birth of the star during the most poorly understood stage of cooling. For insight into the cooling curve of a quark star we follow the work in two papers. The initial cooling of the star is described in \citet{Haensel:1991um}, which focuses on neutrino diffusion through the star and thermal cooling. The star then cools until the neutrinos can escape the quark star and the cooling moves into a purely radiative regime as discussed in \citet{Page:2002bj}.  The part of the cooling curve between these two well defined mechanisms constitutes the translucent regime.

The quark direct Urca processes, which are weak interactions, create both the neutrinos responsible for cooling the star and the parity that drives the current. The topological current only flows when the mean free path of the electron with respect to the weak interaction is larger than the neutron star.  As they take part in the same processes, this is also when neutrinos start to escape the star. The mean free path of the weak interaction for an electron in quark matter has been calculated in \citet{Charbonneau:2009ax} and is
\begin{eqnarray}
\label{eq:l_quark}
\ell_e = 1.2 \times 10^{10} \, \left(\frac{T_\textrm{core}}{10^9 \textrm{ K}}\right)^{-5} \textrm{ km}\,. 
\end{eqnarray}
Equation (\ref{eq:l_quark}) tells us that for a star of radius $R\sim10$ km the current starts to flow at $T_\textrm{core} \sim 7\times10^{10}$~K. At this point on the cooling curve the temperature changes on the scale of seconds.     

\begin{figure} 
\begin{center}
 \includegraphics[width=84mm]{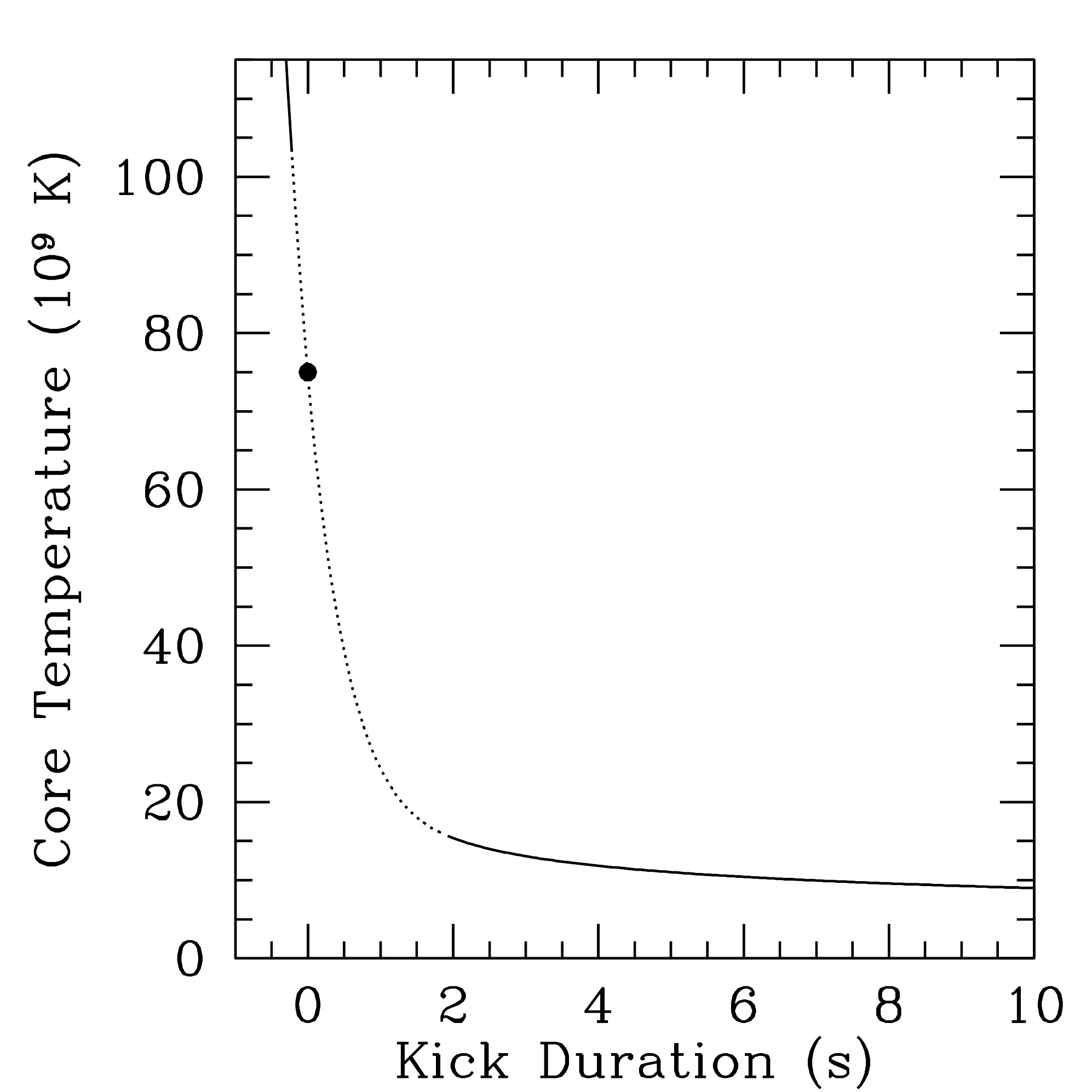}
  \caption{The dashed line indicates the translucent part of the cooling curve, modelled by exponential decay. The curve before the patch is taken from \citet{Haensel:1991um} and the curve after the patch is from  \citet{Page:2002bj}. The black dot marks the start of the kick at $t=0$. }
  \label{figure_cooling}
\end{center}  
\end{figure}

It is not clear how to model the short but critical cooling period that occurs in the translucent regime, as it is a very complicated process, but we assume it must make a smooth transition from the primarily thermal cooling of the core found in \citet{Haensel:1991um} to the neutrino cooling found in \citet{Page:2002bj}. Because we are interested in an estimate of the kick and not the exact details of the cooling we will interpolate between these two cooling curves by assuming the transition is smooth and exponential. We will use the cooling curve from Figure 2a of \citet{Haensel:1991um} that models the cooling of the core of the star. We then piece it together using an exponential decay to cool the star to $10^{10.2}$ K and match the initialization of the cooling curve given in Figure 2 of \citet{Page:2002bj}. By requiring that the first derivatives are equal it is possible to find a unique exponential decay that patches these cooling curves together. Figure \ref{figure_cooling} plots this patched region that describes the translucent region of the cooling curve.

It is possible that the cooling in the translucent regime is quite different from the exponential patch we propose. The severity of the exponential cooling provides us with a reasonable guess for the lower bound of the kick. Slower cooling in the translucent regime would result in larger kicks.

\section{Estimating the size of the kick}

The topological current runs along the magnetic field of the star and consists of electrons created in the core of the star travelling to the surface. Once at the surface these electrons transfer net momentum to the star. We assume that due to the thin crust of the quark star all of the electron's momentum is able to be transferred. The electrons may be physically ejected from the star or the momentum may be transferred by some radiative process such as bremsstrahlung.  The size of the kick is directly related to the strength of the current, which is related to the creation rate of left-handed electrons in the quark star and the population of electrons in the lowest Landau level. The calculations regarding the strength of the current in this paragraph are detailed in \citet{Charbonneau:2009ax}. We provide only a brief illustrative overview here.  The direct Urca processes for quarks create electrons at a rate per unit volume
\begin{eqnarray}
\label{eq:rate_quark}
\frac{w}{\Omega} &=&  \frac{G_F^2}{\pi^6}\,\alpha_s \,k_e\, k_q^2\, T^5 \,I \\
&=& 4.4 \times 10^{31}\, \left(\frac{n_b}{n_0}\right)\, \left(\frac{T_\textrm{core}}{10^9 \textrm{ K}}\right)^{5}  \textrm{ s}^{-1} \textrm{ cm}^{-3} \,,
\end{eqnarray}
where $\alpha_s = g^2/4\pi \approx 0.4$ and $I\approx20.6$ is a dimensionless integral. The helicity of the electrons $\langle \Lambda \rangle = -0.84$ and the fraction of electrons in the lowest Landau level are contained in the parameter,
\begin{eqnarray}
\label{P}
P_{\textrm{asym}}(\mu_e, T, B)  &\simeq&  - \langle \Lambda \rangle\cdot  \left(\frac{m_e^2}{k_e^2}\frac{B}{B_{\textrm{c}}}\right) \\
&\sim& 2\cdot10^{-5} \left(\frac{B}{B_{\textrm{c}}}\right)\left(\frac{n}{n_0}\right)^{-2/3},
\end{eqnarray}
The total number current for electrons reaching the surface of the star is given by,
\begin{eqnarray}
\label{eq:current_quark}
\langle j \rangle &=&   P_{\textrm{asym}}\, \frac{w}{\Omega}\, V_{\textrm{star}} \\ &=& 2.6 \times 10^{26}\left( \frac{73 \textrm{ MeV}}{k_e} \right) \left( \frac{B}{10\, B_\textrm{c}} \right) \nonumber\\ && \times\left(\frac{T_\textrm{core}}{10^9 \textrm{ K}}\right)^{5} \left(\frac{n_\textrm{b}}{n_0}\right)^{1/3} \textrm{MeV}\,, 
\end{eqnarray}
where $B_\textrm{c} = 4.4\times10^{13}$ G is the critical magnetic field, $T_\textrm{core}$ is the core temperature of the star, and $n_0$ is nuclear density. The typical density for quark matter is $n_\textrm{b} \sim 10\, n_0$ but could easily be higher. 

Because the current originates in the core of the star, we are interested in the strength of the internal magnetic field. The dipolar component of the Vela pulsar at its surface is about  $B_\text{s} \sim 3.5\times10^{12}$~G  \citep{Abney:1996kn} but the field in the bulk of the star is likely much stronger based on virial theorem arguments in \citet{Lai:1991}.  The scalar virial theorem for the star states that 
\begin{eqnarray}
2T + W + 3 \Pi + \mathcal{M} = 0,
\end{eqnarray}
where $T$ is the total rotational kinetic energy, $\Pi$ is the internal energy, $W$ is the gravitational energy, and $\mathcal{M}$ is the magnetic energy. Because the kinetic energies are positive, the magnetic field energy can be at most equal to the gravitational energy, $(4\pi R^3/3)( B_\textrm{max}^2/8\pi) \sim GM^2/R$, which yield possible core fields of $B_\textrm{max} \sim10^{18}$ G. This is an extremely large field and is unlikely as it is a strict upper bound. Arguments by  \citet{Abney:1996kn}  based on glitch timing of the Vela pulsar may put minimums on the core field depending on whether the protons are superconductors or not. As pointed out  in \citet{PhysRevLett.38.1425}, currents along type-II vortices  are likely to destroy superconductivity. \cite{Charbonneau:2007db} also point out that the presence of topological currents could similarly destroy superconductivity in neutron stars, thus the non-superfluid estimate for the internal field from \citet{Abney:1996kn} is appropriate. In very dense matter they find minimum interior fields of $B_\textrm{min} \sim10^{14}$ G.  Based on this we choose a value of the core magnetic field to be $B_\textrm{core}=10\,B_\textrm{c}$. 

The current, equation~(\ref{eq:current_quark}), gives a number rate of electrons reaching the surface of the star. To get the momentum transferred per second we simply multiply by the momentum each electron carries, $k_e$, which is large due to the degeneracy and is given by equation~(\ref{eqn:k_e}). The acceleration for a $1.4 M_\odot$ star is then,
\begin{eqnarray}
a = \frac{k_e\ \langle j(T_\textrm{core}) \rangle}{1.4M_{\odot}}.
\end{eqnarray}
We see that all the factors of $k_e$ cancel because $\langle j \rangle$ contains a factor of $k_e^{-1}$. The kick is independent of the Fermi momentum of the electrons, but as we will discuss later this does not mean the Fermi momentum is unimportant.

\begin{figure}
  \begin{center}
  \includegraphics[width=8cm]{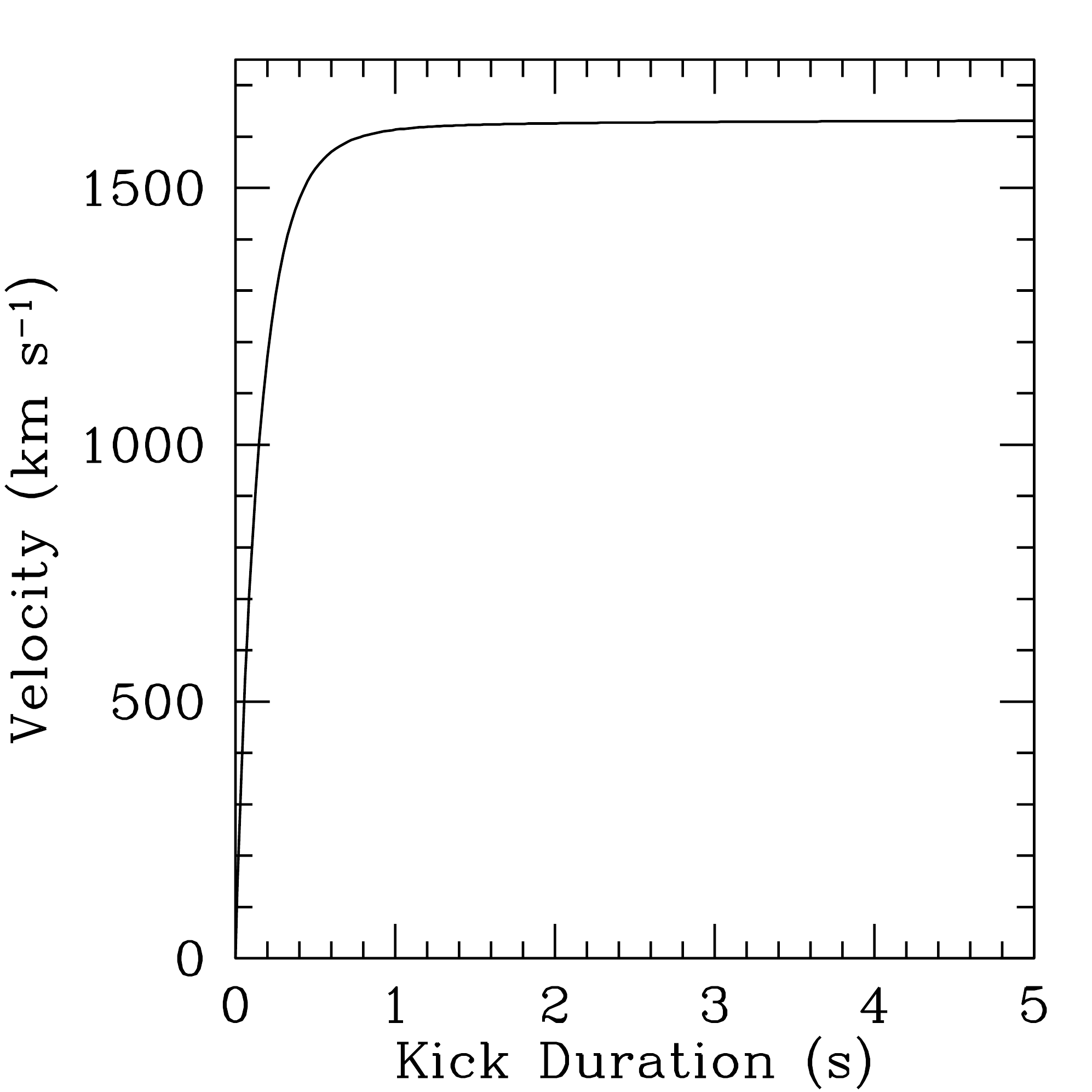}
  \caption{Time evolution of the kick for an internal magnetic field $B=10\, B_\textrm{c}$.}
  \label{figure_kick}
  \end{center}
\end{figure}

As seen in Figure \ref{figure_kick}, the star quickly reaches a speed of
\begin{eqnarray}
v_{\textrm{max}} \sim 1600 \textrm{ km$\,$s$^{-1}$}\,,
\end{eqnarray}
which is big enough to account for the large kicks seen in many pulsars. This result relies on the internal magnetic field of the star being two orders of magnitude larger than what we observe on the surface of most pulsars. We have also neglected the dissipation of electrons in the very thin crust of the quark star. As plotted, the entire kick seems to happen very quickly, but the current keeps running throughout the star's life. With a large internal magnetic field the mechanism can account for kicks seen in young pulsars such as Vela.  But because the kick is constantly running, pulsars with smaller internal magnetic fields will eventually attain very large speeds very late in life.

\break
\section{How are topological kicks different than neutrino kicks?}
\label{difference}
It is important to understand the differences between neutrino kicks and topological kicks because superficially they seem very similar. In both cases a particle is created through the weak interaction with a favoured helicity (left-handed)\footnote{In a dense star the electrons are created with an average helicity $\langle \Lambda \rangle = -0.84$, but for simplicity we will assume in this section that they are all left-handed.} and the large magnetic field forces the electrons into Landau levels. The lowest Landau level only accepts a spin down electron, whereas all the rest allow both spin up and spin down states.  Combining the preferred helicity of the particles with the spin state asymmetry in the lowest Landau level means that the neutrinos and electrons now travel with a preferred direction within the star. If the star is cool enough, these particles can escape the star without decaying.  The neutrino does not participate in QED interactions and the electrons propagate due to the current, which is allowed because parity is a symmetry of QED. The particles that exit the star provide the linear momentum required to push the star.       

The electrons and neutrinos that contribute towards the kick are created at the same rate $w$, given by equation (\ref{eq:rate_quark}). We are only interested in interactions that create particles in the lowest Landau level so this rate is suppressed by a factor to account for the population of the lowest Landau level. The momentum transferred to provide the kick is simply the momentum per particle that escapes the star.  On the surface of the star, where the chemical potential is low, Boltzmann statistics are used to model the occupation of the lowest Landau level. This is what is seen in many neutrino kick models. This is because neutrinos require such hot temperatures to produce kicks that only a thin shell near the edge of the star is considered to contribute to the kick. In the core of the star, where the electrons are very degenerate, we must use Fermi statistics. This predicts an occupation of
\begin{eqnarray}
n_\textrm{L}\sim\left(\frac{m_e^2}{k_e^2}\frac{B}{B_{\textrm{c}}}\right)
 \end{eqnarray}
in the lowest Landau level in the interior of the star. 

At core temperatures below $7\times10^{10}$~K the neutrinos can escape the star and the topological current is active. The flux of particles contributing to both the electron kick and the neutrino kick is about the same $\sim n_\textrm{L}\,w$. The difference between the two mechanisms comes from the momentum that the relevant particle carries. The neutrinos are created thermally and the typical momentum of a neutrino is equal to the temperature of the star $T$. The momentum of the electrons comes from the large chemical potential, $\mu_e\sim 150$ MeV. The momentum transfer per unit time for neutrinos\footnote{Notice energy per unit time has the $T^6$ dependance we expect for a calculation involving neutrino  luminosity.}  is $F_\nu \sim T n_\textrm{L} w$ and for electrons is $F_e \sim \mu_e n_\textrm{L} w$. When the kick starts the star has a temperature of only $T\sim 1$ MeV. The electron kick is stronger than the neutrino kick by a factor of
\begin{eqnarray}
\frac{F_e}{F_\nu} \sim  \frac{\mu_e } {T}\,.
 \end{eqnarray}
Initially, when the star is very hot, the electron kick is two orders of magnitude stronger than the neutrino kick. Furthermore, as the star cools the neutrino kick gets even weaker, while the electrons continue to have a momentum dictated by their chemical potential. This why electrons generate larger kicks than neutrinos.

\section{How much does the current affect the cooling of the star?}

At the beginning of the star's life the energy from the kick does not contribute to the cooling of the star, but later in life the current could become the dominant cooling mechanism. This may account for the colder than normal temperatures seen in the Vela pulsar.

The reason is that only a small fraction of the electrons created in the neutron star actually escape due to the low occupation of the lowest Landau levels. As discussed earlier the estimate for the neutrino kick has a similar suppression factor, but all of the neutrinos actually escape the star, even those that do not contribute toward a neutrino kick. The electrons only propagate because the asymmetry in the lowest Landau level allows the helicity states to propagate out of the star.  Those electrons that do not contribute toward the kick are trapped inside the star. The helicity states that reach the surface due to the current are the only electrons that contribute to the cooling of the star. The neutrinos cool the star with a luminosity $L_\nu \sim T w$ where the electrons cool the star with an energy current (luminosity) of $L_e \sim k_e n_\textrm{L} w$. The ratio of electron cooling to neutrino cooling is
\begin{eqnarray}
\frac{L_e}{L_\nu}\sim\frac{\mu_e n_L} {T}\,.
 \end{eqnarray}
At the star's birth the electrons cool the star at about 1/100 the rate of neutrino cooling. As the star cools, eventually $L_e/L_\nu > 1$ and the current becomes the dominant cooling mechanism of the star. This transition occurs at $T_\textrm{t}\sim 0.1 \textrm{ MeV} \sim10^9$ K. The current is an additional cooling mechanism to consider in stars that have cooled below $10^9$ K. This may explain the unusually cold surface temperature of $(7.85\pm0.25)10^5$ K that is observed \citep{Page:1996xw} in the Vela pulsar.

\section{Summary}
We have found that the topological currents introduced in \citet{Charbonneau:2009ax} can generate kicks greater than 1000~km$\,$s$^{-1}$. This may explain large pulsar kicks that are otherwise unattainable using traditional kick mechanisms. Topological kicks occur only if the electrons that make up the topological current can transfer their momentum through the star's crust. Because neutron stars have thick crusts, it is unlikely that they will be kicked due to this mechanism, but rather by mechanisms that generate weak kicks. In contrast, bare quark stars have very thin crusts that may allow the electrons to escape. Because of this distinction we conjecture that many of the fastest pulsars are quark stars.

In calculating the kick velocity we assumed that the cooling curve of the quarks star must make a smooth transition from the primarily thermal cooling of the core found in \citet{Haensel:1991um} to the neutrino cooling found in \citet{Page:2002bj}. This patched cooling curve allowed us to integrate over time the momentum transferred from the topological current and obtain the strength of the kick. We have also demonstrated that the energy lost due to the current does not change the cooling of the star during the kick. But, after the star cools below $10^9$~K the topological current becomes the dominant cooling mechanism rather than neutrino emission.

The magnitude of the kick also depends on the strength of the magnetic field inside the star.  The virial theorem puts a limit on the internal field of the star that is much higher than the surface field. To generate the kick we used a field strength of $B_\textrm{core} = 10\,B_\textrm{c}$. This is a reasonable estimate for pulsars that have a surface magnetic field of $B_\textrm{s} \sim 10^{12}$ G such as the Vela pulsar.

\section*{ACKNOWLEDGMENTS} 

This work was supported by the National Science and Engineering Council of Canada. The authors thank Jean-Fran\c{c}ois Caron and George Hicks for their useful comments and corrections.

\bibliographystyle{mn2e}

\bsp

\label{lastpage}

\end{document}